# Two-photon interference from as-grown InAsP/InP quantum dots under detuned excitation

*Maja Wasiluk*, Anna Musiał, Paweł Mrowiński, Johann Peter Reithmaier, Mohamed Benyoucef, Wojciech Rudno-Rudziński*

M. Wasiluk, A. Musiał, P. Mrowiński, W. Rudno-Rudziński
Department of Experimental Physics, Faculty of Fundamental Problems of Technology, Wrocław University of Science and Technology, Wyb. Wyspiańskiego 27, 50-370 Wrocław, Poland
E-mail: maja.wasiluk@pwr.edu.pl

J. P. Reithmaier, M. Benyoucef
Institute of Nanostructure Technologies and Analytics, CINSaT, University of Kassel, Heinrich-Plett-Str. 40, 34132 Kassel, Germany

Funding: This work was co-financed by the Ministry of Education and Science, Republic of Poland within the ”Perły Nauki” project, grant no. PN/01/0117/2022. This work was also financially supported by the Bundesministerium für Forschung, Technologie und Raumfahrt-BMFTR in the frame of the project QR.N (16KIS2204), the DFG-Heisenberg grant (505496601), and the DFG project (430159793).

Keywords: III-V semiconductor, InAsP/ InP quantum dots, two-photon interference, indistinguishable single photons, telecom spectral range

**In this study, we investigate as-grown InAsP/InP quantum dots emitting in the third telecommunication window. The investigated quantum dots are embedded directly in an aluminium-free InP matrix without additional barrier material. A large excitation–emission detuning of 32 meV enables efficient suppression of scattered laser light while retaining several advantages of near-resonant excitation. The single-photon nature of the emission is confirmed by a Hanbury Brown and Twiss experiment, yielding a raw**

**second-order autocorrelation value of $g^{(2)}_{\mathrm{raw}}(0) = 0.076(6)$. A Hong–Ou–Mandel measurement is used to determine the degree of indistinguishability of single photons and reveals as-measured visibilities of $V = 0.038(8)$ and $V = 0.108(18)$ for excitation pulse separations of 13.1 ns and 5.3 ns, respectively. We observe finite photon indistinguishability in as-grown InAsP/InP quantum dots under detuned quasi-resonant excitation, without deliberate Purcell enhancement.**

## 1. Introduction

Near-ideal sources of indistinguishable single photons are an essential resource for many photonic quantum technologies. They are necessary for Bell state measurements used in quantum repeaters[1,2], as well as for linear optical quantum computing[3–5], boson sampling[6], and the generation of photonic cluster states[7]. Both low probability of multiphoton events and high visibility in two-photon interference are crucial for quantum information processing; nevertheless, for practical implementation, the source needs to meet several other requirements[8]. A practical single-photon source should provide high brightness to enable efficient, fast data transmission. This performance is typically quantified by the probability of delivering a single photon to the device output per excitation pulse, ideally with a GHz repetition rate[8]. Furthermore, seamless integration into an all-fiber photonic platform with efficient end-to-end fiber coupling is highly desirable [9]. Unfortunately, in all-fiber applications, polarization-based filtering is often compromised due to fiber birefringence[10,11] and the lack of stable polarization maintenance.

To overcome the challenges of stray laser light rejection associated with strictly resonant fluorescence, advanced excitation schemes, such as longitudinal acoustic (LA) and optical (LO) phonon-assisted or swing-up population have been explored.[12–15] LO-phonon-assisted excitation is particularly appealing, as it allows for straightforward laser rejection due to the large spectral detuning from the transition of interest. Great progress in the design and realization of fiber-coupled devices brought us closer to achieving the practical advantages[16–18]; nevertheless, no platform has yet simultaneously demonstrated near-unity single-photon purity, high indistinguishability, high brightness at the fiber output, and emission in the telecom C-band.

Over the recent decades, advancements in epitaxial growth have led to better nanostructure quality, while more sophisticated fabrication techniques, such as deterministic integration of quantum dots (QDs) into photonic microstructures,[19,20] have been developed. Furthermore, the transition from non-resonant to other excitation schemes has increased the visibility of quantum interference.[12,21,22] Alongside the low multiphoton emission probability achieved for quantum dots (QDs) emitting in the telecom C-band, substantial progress has been made in the two-photon interference (TPI). InAs/InP QDs grown by metalorganic vapour phase epitaxy (MOVPE) using droplet epitaxy exhibited peak TPI visibilities of 71.3% under non-resonant continuous-wave (cw) excitation[23] and 77% under resonant cw excitation[24]. Under pulsed excitation, integrated TPI visibilities of approximately 35% were reported for MOVPE-grown InAs/InP QDs[25] and In(Ga)As QDs grown by MOVPE[21], followed by raw values of 71.9% and 91.7% for MBE-grown InAs/InAlGaAs/InP QDs.[12,26]

Here, we investigate the two-photon interference from molecular beam epitaxy (MBE)-grown InAsP/InP quantum dots non-deterministically embedded in mesa structures for measurements repeatability. The studied emitters are used in their as-grown form. A distinctive feature of this material system is that the InAsP quantum dots are embedded directly in an aluminium-free InP matrix, without additional barrier material. For the quantum-optical measurements, the excitation was into a higher energy state within a QD, and its power was set to the saturation level, corresponding to the operating regime most relevant for practical single-photon-source applications, where high photon generation rates are required. A photon extraction efficiency of approximately 13.3% into an objective with a numerical aperture of 0.4 was previously reported for this mesa–DBR platform.[27]

## 2. Results and Discussion

### 2.1. Optical characterization

Quasi-resonant excitation into higher-energy states of the QD is a convenient method for carrier generation, as it simplifies the spectral filtering of the excitation laser. To determine the suitable excitation wavelength, we first performed the photoluminescence excitation (PLE) measurement. Emission intensity as a function of the excitation energy is shown in **Figure 1(a)**. We focused our study on the transition highlighted between the grey lines, which exhibits the highest emission intensity over a broad range of excitation energies. The broad excitation resonance provides efficient excitation over a relatively wide energy range, reducing the sensitivity of the excitation efficiency to small fluctuations in the laser wavelength. The right panel of Figure 1(a) shows the PLE spectrum for the studied line, where two prominent resonances are observed at the energy detuning of $\Delta E = 32$ meV and $\Delta E = 38$ meV. For quantum optical measurements, the resonance at $\Delta E = 32$ meV was selected due to its higher intensity and a flatter spectral profile in the vicinity of the investigated transition (**Figure 1(b)**).

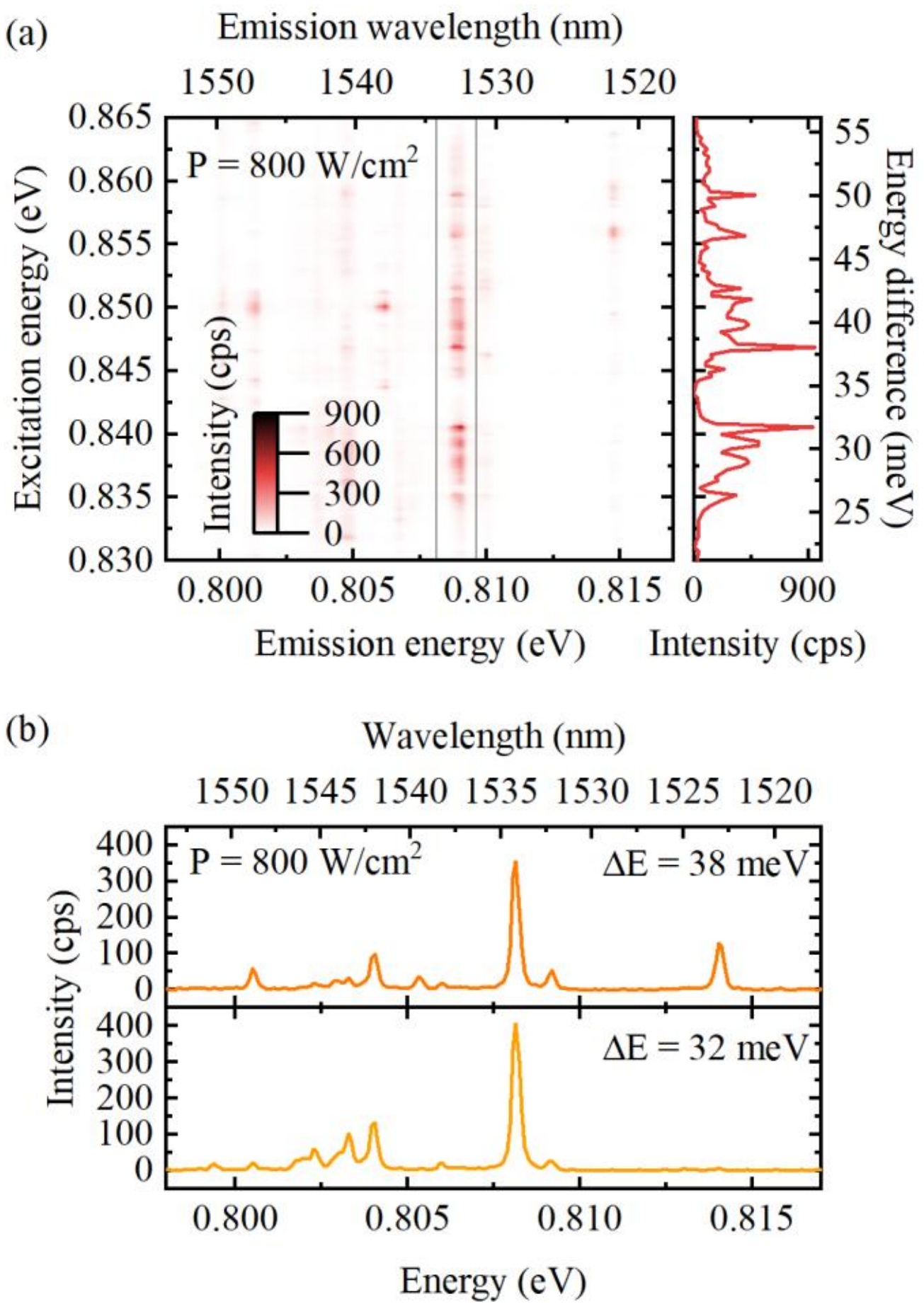

**Figure 1.** (a) PLE spectra of studied QD with investigated line (within grey vertical lines) scanned with 1 nm of laser wavelength step (corresponding to ~280-300 μeV). Right panel, intensity of investigated line as a function of energy difference between excitation and emission. Two strong resonances visible for energy differences of 32 meV and 38 meV. (b) PL spectra for two excitation energy detunings: 38 meV (upper panel) and 32 meV (bottom panel) above the emission energy.

The photoluminescence (PL) spectrum under the non-resonant excitation (at the saturation power density of 160 W/cm$^2$) is shown in **Figure 2(a)**. While the excitation power density is almost 6 times lower than the maximal value used under quasi-resonant excitation, the investigated emission line did not reach saturation in the PLE measurement. This is attributed to the much smaller absorption cross-section of a single QD compared to the bulk substrate and barrier materials and the limited laser power available on the sample. Under non-resonant excitation, an emission background and many neighboring emission lines are observed, particularly in the close spectral proximity of the studied line (indicated by the black arrow).

Inset of Figure 2(a) shows the integrated intensity of the Gaussian fit as a function of excitation power density. The data follows a power-law dependence, with a slope of 1.9. In a typical QD system (simple two-level system), a slope of 1.0 is characteristic of neutral excitons, whereas a slope near 2.0 indicates the recombination of a biexciton in a neutral or charged state.[28]

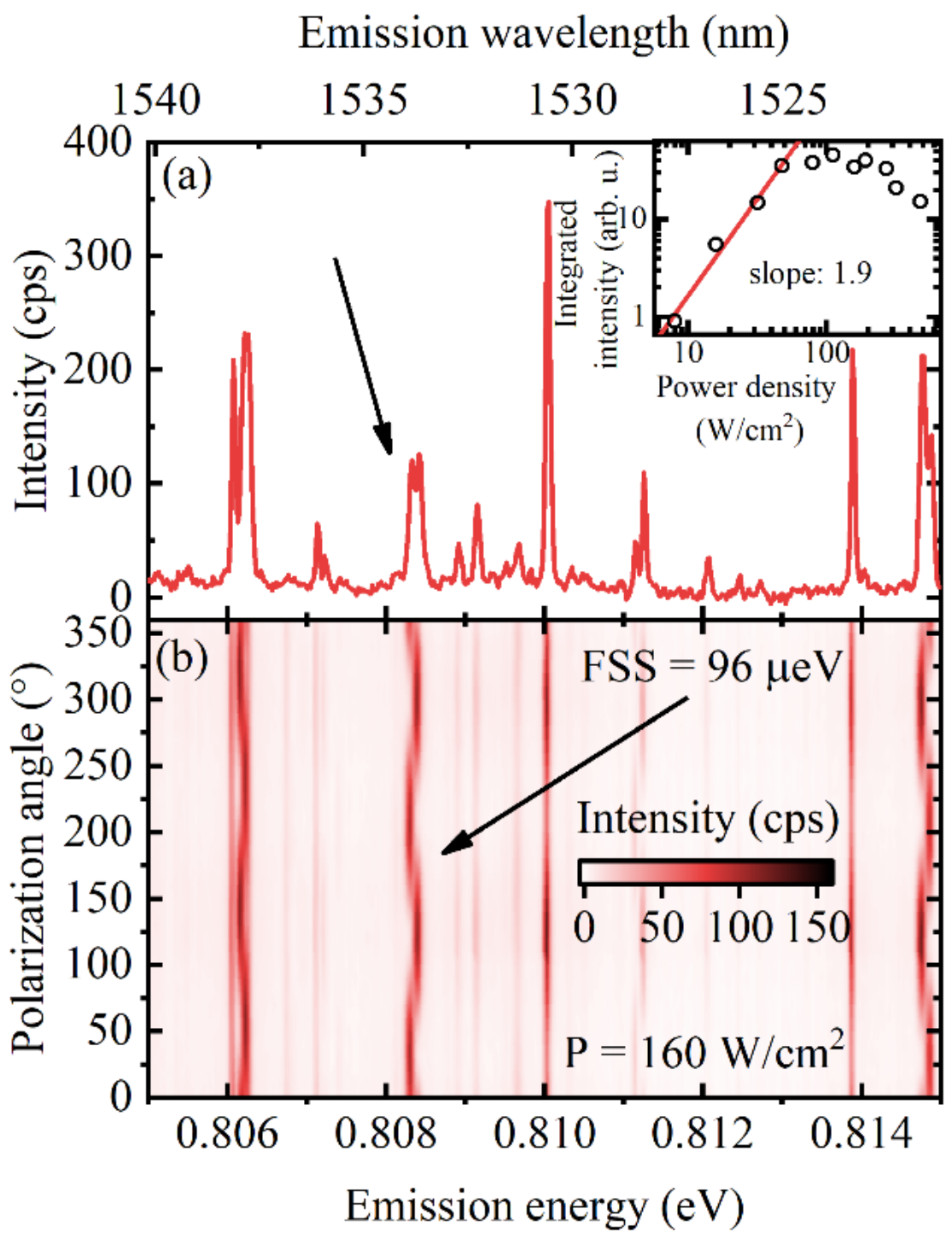

**Figure 2.** (a) Non-resonant excitation spectrum of studied QD with the emission line under investigation marked by an arrow. Selected excitation power density of 160 W/cm$^2$ corresponds to near-saturation power. Inset: Emission intensity as a function of excitation power density, fitted with a power-law dependence with an exponent of 1.9. (b) Polarization-resolved map of photoluminescence under non-resonant excitation at the power density of 160 W/cm$^2$, revealing FSS of studied line of 96 μeV.

The polarization-resolved PL studies, presented in **Figure 2(b)**, reveal a distinct fine-structure splitting (FSS) of the emission lines, a feature typically arising from the exchange interaction in asymmetric nanostructures. Clear splitting of the emission energy as a function of the detection angle provides a characteristic signature of orthogonally linearly polarized doublet. For the investigated line, we determine an FSS of 96 μeV (experimental spectral resolution is approx. 20 μeV). Given the quadratic power dependence, this transition is attributed to the recombination of a biexciton state.

Under these excitation conditions, the emission linewidth measured in one linear polarization is 84 μeV, which is nearly two orders of magnitude broader than the transform-limited linewidth (< 1 μeV, based on a measured lifetime (~1 ns)). This significant broadening suggests strong spectral diffusion, likely induced by fluctuating charge traps in the QD's solid-state environment.

### 2.2. Multi-photon emission probability

Excitation into a higher energy state of the QD resulted in background reduction (Figure 1), and linewidth narrowing to 62 μeV, suggesting a decreased density of fluctuating carriers in the emitter's vicinity, likely due to a lower overall number of photogenerated carriers in this excitation scheme. For subsequent quantum optical experiments, the excitation power was set close to the saturation power to approximate the operating regime of a high-brightness source, as discussed in the introduction. The experiment utilized a pulsed laser with a repetition rate of 76 MHz, corresponding to a temporal pulse separation of 13.1 ns.

The Hanbury Brown and Twiss (HBT) measurement confirmed a low probability of multiphoton emission events. **Figure 3** presents the second-order correlation function $g^{(2)}(\tau)$ over a 190 ns range, showing a pronounced reduction of coincidences at the zero time delay. Histograms are normalized to the Poissonian limit where $g^{(2)}(\tau \to \infty) = 1$ and shown without any background subtraction or blinking correction. The observed exponential envelope in the histogram indicates the blinking dynamics, affecting the emission state with the characteristic time $\tau_{\text{blink}} = 81(2)$ ns. By fitting the correlation histogram with double exponential decay functions, including the blinking component,[29–31] we evaluated the multiphoton emission probability to be $g^{(2)}_{\text{fit}}(0) = 0.0091(88)$. However, to provide a more conservative and insightful metric, we calculated the ratio of integrated raw data counts within the excitation pulse period. By comparing the area of the central peak to the mean of the 8 peaks for $\tau \to \infty$, where $g^{(2)}(\tau) \to 1$, we obtained a value of $g^{(2)}_{\text{raw}}(0) = 0.076(6)$. Finally, the radiative lifetime extracted from the decay dynamics was found to be $\tau_{\text{dec}} = 1.165(7)$ ns.

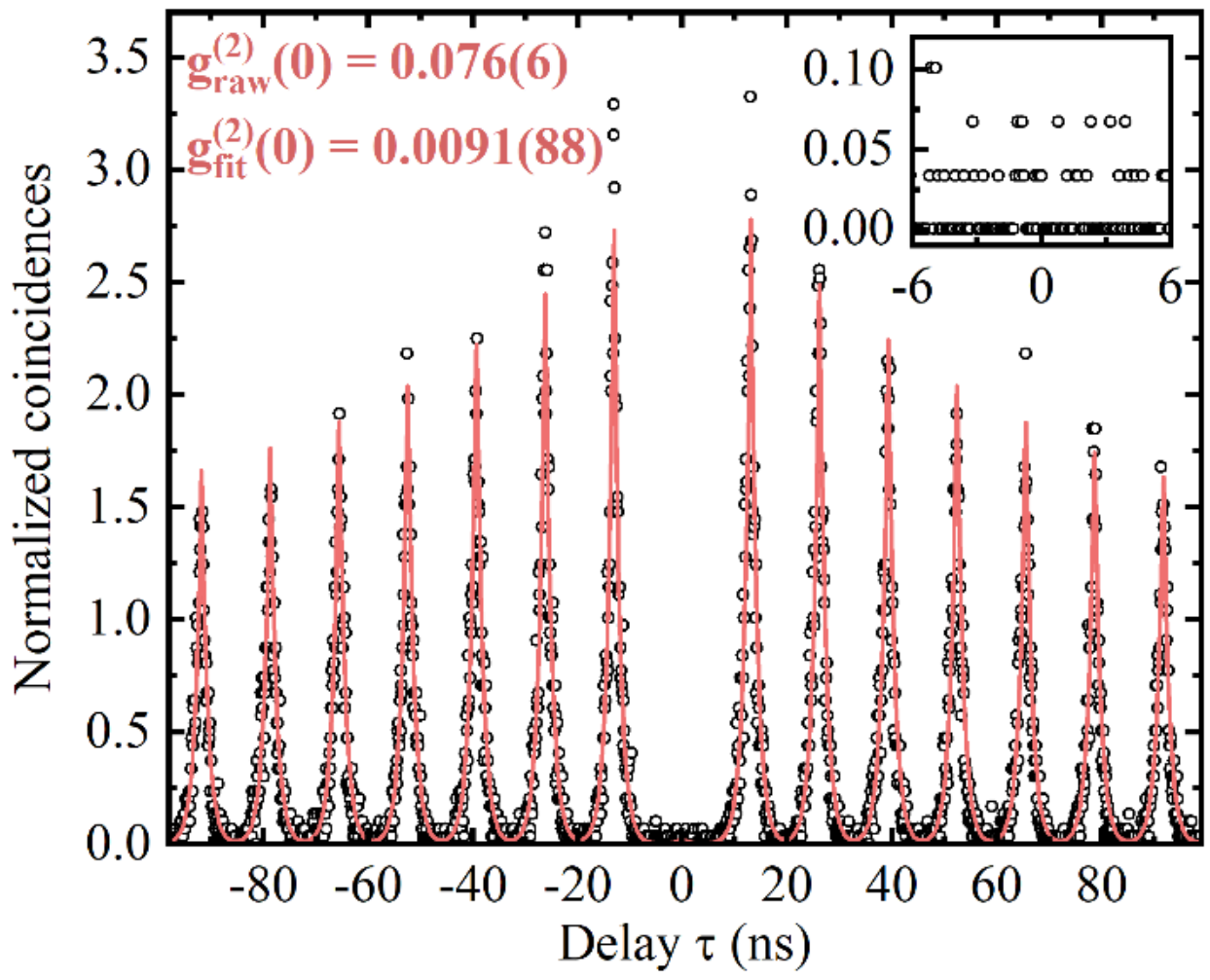

**Figure 3.** Autocorrelation $g^{(2)}$ function histogram for saturation power $P_{sat}$ with strong reduction of coincidences visible for central peak around zero time delay. Inset shows the same data for narrow range of arrival delays within central peak.

## 2.3. Two-photon interference

### *2.3.1. Pulse separation of 13.1 ns*

To evaluate the degree of indistinguishability of the emitted single photons, Hong-Ou-Mandel (HOM) interferometry experiment was performed under the same excitation conditions as the second-order autocorrelation measurements. The resulting coincidence histogram, fitted with double-exponential decay functions (including the blinking effect), is presented in **Figure 4**. A broad-range overview of the detection events (Figure 4(a)) reveals a bunching indicative of source blinking, with an extracted characteristic time $\tau_{blink} = 45(1)$ ns. Although the pulse separation is relatively large, the slight overlapping of adjacent peaks is visible. The extracted decay time ($\tau_{dec} = 1.35(1)$ ns) is slightly longer than that obtained from the HBT measurements. This difference most likely arises from the fitting procedure, residual overlap of neighboring coincidence peaks, and finite detector timing resolution, rather than from a change in the emitter lifetime. For quantitative analysis, the data were normalized to the Poissonian limit. A closer look at the central peak (Figure 4(b)) reveals a suppression of coincidences, with the zero-delay visibility of $V_{fit} = 0.544(63)$ and the coherence time $T_2 = 114(27)$ ps, extracted from the dip observed in the central peak.

To quantify the experimentally observed indistinguishability over the full pulse period, we analyzed the whole central peak, within the time scale of pulse separation. From the lack of representative results for the perfectly distinguishable photons, we simulated the central peak shape based on the height of the surrounding peaks for $\tau \neq 0$ (grey solid line in Figure 4(c)). Based on this, we estimated the visibility of $V = 0.038(8)$ for the 13.1 ns laser pulse separation.

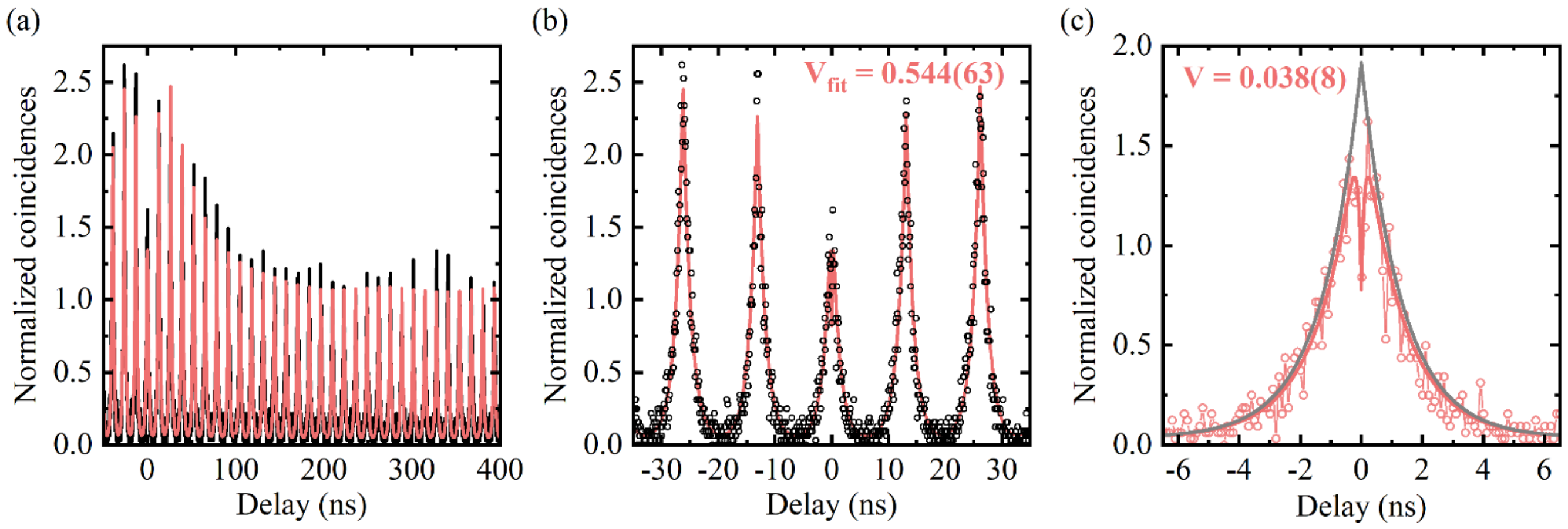

**Figure 4.** (a) Normalized HOM histogram data (black line) for a wide range of delay times, showing blinking with characteristic time $\tau_{\text{blink}} = 45(1)$ ns. (b) Zoomed-in histogram showing middle peak, with 2 surrounding peaks from each side. Fitting (red line) resulted in HOM visibility determination $V_{\text{fit}} = 0.544(63)$. (c) Coincidences within middle peak of 13.1 ns total range (red points with line) with fitting (red line) and predicted peak for completely distinguishable photons (grey line), resulting in visibility estimation of $V = 0.038(8)$.

*2.3.2. Pulse separation of 5.3 ns*

To further investigate the temporal dynamics of the photon coherence and evaluate the impact of spectral diffusion on shorter timescales, the pulse separation was reduced to 5.3 ns by adding a Mach-Zehnder interferometer in the excitation path. By decreasing the time interval between consecutive emission events, the measurement becomes less sensitive to slow fluctuations of the solid-state environment. The resulting HOM interference histogram is presented in **Figure 5**, manifesting blinking ($\tau_{\text{blink}} = 56(1)$ ns) of the source, consistent with the previous measurement. In comparison to the measurements performed with a delay of 13.1 ns, the stronger suppression of coincidences at zero-time delay indicates a higher degree of photon indistinguishability. The shorter temporal separation reduces the impact of slow dephasing processes, allowing the emitted photons to retain a greater overlap of their quantum states. Using the same double-exponential fitting procedure, we extracted the zero-delay visibility of $V_{\text{fit}} = 0.70(10)$ and the coherence time $T_2 = 177(48)$ ps (Figure 5(b)). Short laser pulse separation and long decay time resulted in a strong overlapping of the peaks. To determine the value of visibility for the time-scale of the whole central peak, we extracted the individual peaks originating from all possible interference paths in both interferometers. The neighboring peaks were used to reconstruct the central peak expected for distinguishable photons, taking into account the same blinking-induced intensity fluctuations as for the measured central peak. Using $V = 1 - A_{\parallel}/A_{\perp}$, where $A_{\parallel}$ and $A_{\perp}$ are the areas of the fitted and reconstructed central contributions integrated over a 5.3 ns window centred at zero delay, we determined the visibility to be $V = 0.108(18)$. Integrating the full fitted coincidence histogram and its simulated distinguishable-photon reference over a 5.3 ns window, without separating the individual peaks, yielded a visibility of $V=0.061(11)$. The dependence of the extracted visibility on the integration window is shown in the Supplementary Material.

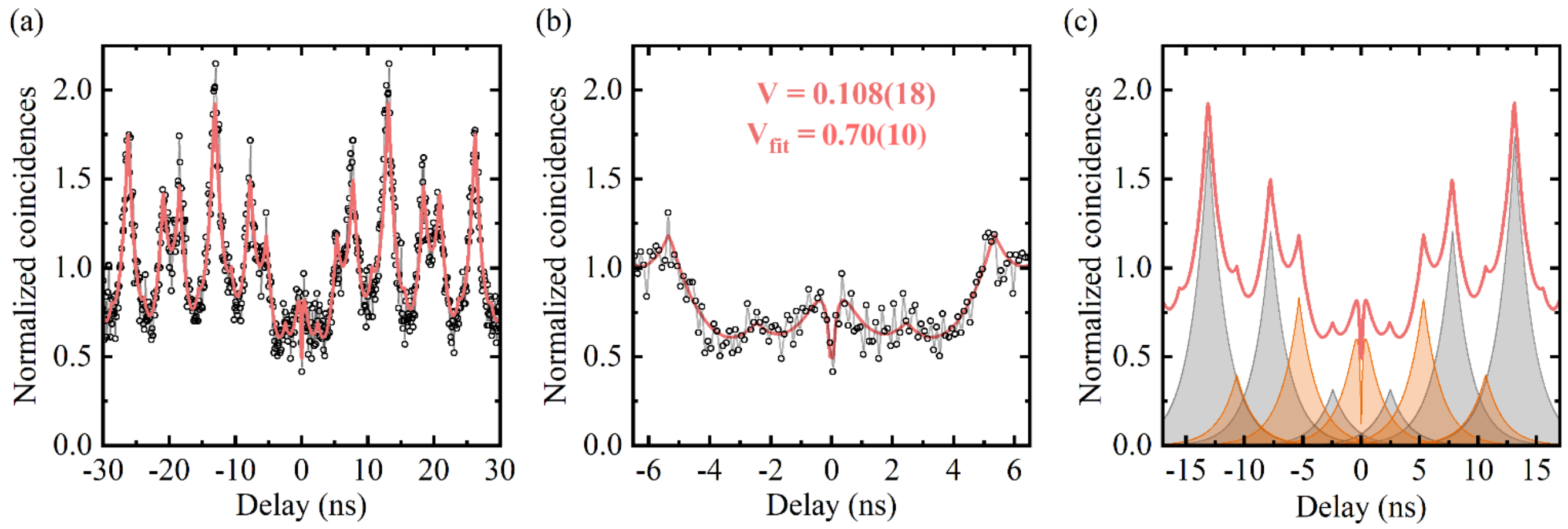

**Figure 5.** (a) Normalized HOM histogram data (black symbols and line) for a wide range of delay time showing blinking with the characteristic time $\tau_{\text{blink}} = 56(1)$ ns; fitted double exponential decay function (red line). (b) Same histogram for narrow range of delay time for closer look at central peak with reduced coincidences. (c) Fitting function (red line) with extracted parts originating from central 5-peaks cluster (orange shaded area) and neighboring peaks (grey shaded area).

**Table 1** summarizes the parameters extracted from the analysis of the coincidence histograms obtained for the $g^{(2)}$ function and the two-photon interference measurement for laser pulse separations of 13.1 ns and 5.3 ns. The demonstrated low value of $g^{(2)}(0)$, together with the non-zero degree of photon indistinguishability, highlights the potential of the investigated as-grown quantum dots as quantum light sources for emerging quantum technologies. The extracted decay times vary between experiments. The increase of $\tau_{\text{dec}}$ from HBT to HOM measurements may likely arise from the fitting procedure. The fitting model assumes a monoexponential PL decay, whereas time-resolved PL measurements reveal a non-monoexponential decay, particularly at higher excitation powers (see the Supporting Information for a detailed analysis). This simplification may also contribute to the observed increase in the extracted decay times. The extracted values of $\tau_{\text{blink}}$ also differ between the measurements. This difference is likely related to the substantially different acquisition times, which affect the statistical sampling of the blinking dynamics.

**Table 1.** Parameters obtained from fitting of $g^{(2)}$ and HOM histograms. The precision of the extracted parameters are only related to the fitting procedure and not to the total experimental accuracy.

| Experiment | Pulse separation [ns] | $g^{(2)}_{\text{fit}}(0)$ | $g^{(2)}_{\text{raw}}(0)$ | $V_{\text{fit}}$ | $V$ | $T_2$ [ps] | $\tau_{\text{dec}}$ [ns] | $\tau_{\text{blink}}$ [ns] |
|---|---|---|---|---|---|---|---|---|
| $g^{(2)}$ | 13.1 | 0.0091(88) | 0.076(6) | - | - | - | 1.165(7) | 81(2) |
| HOM | 13.1 | - | - | 0.544(63) | 0.038(8) | 114(27) | 1.35(1) | 45(1) |
| HOM | 5.3 | - | - | 0.70(10) | 0.108(18) | 177(48) | 1.57(3) | 56(1) |

The larger uncertainties associated with the HOM measurements with the 5.3 ns laser pulse separation originate from the substantial overlap of neighbouring peaks, which reduces the reliability of the fitting procedure. Nevertheless, a higher degree of indistinguishability is observed for the shorter pulse separation. The only experimental parameter changed between the two measurements is the temporal separation between the excitation pulses, while the emitter, excitation conditions, and detection scheme remain unchanged. An increase in the HOM visibility for shorter pulse separations has previously been attributed to non-Markovian noise in the literature and a spectral diffusion typical for epitaxial QDs has been identified as the underlying physical mechanism.[35] This observation alone is not sufficient to establish non-Markovian dephasing. In the model of Thoma et al., the TPI visibility is expected to saturate for pulse separations much shorter than the correlation time of the fluctuations. Our measurements at two pulse separations do not establish whether this saturation occurs. However, in these conditions it would not be possible to separate neighboring maxima in the TPI histogram and the analysis would not be reliable in our case (without Purcell enhancement). It was shown that, in the case of Markovian dephasing, the HOM visibility does not depend on the temporal separation between the two excitation pulses, e.g., in the case for phonon-induced dephasing.[36] Since the pulse separations used in the experiment (5.3 ns and 13.1 ns) exceed the extracted optical coherence time $T_2$ by more than an order of magnitude, the observed dependence of the HOM visibility on pulse separation suggests the presence of an additional decoherence mechanism, with a finite correlation time on the nanosecond timescale. This behavior is consistent with slow spectral diffusion[37], which introduces temporal correlations between consecutive emission events and reduces photon indistinguishability for longer pulse separations. The broad emission linewidth observed for the investigated quantum dots indicates significant fluctuations of the transition energy, supporting this interpretation.

**Table 2** compares TPI measurements from quantum dots emitting in the telecom C-band across several material systems, growth methods, and device geometries. The studies cover dots in planar structures, mesas, and circular Bragg grating cavities, with excitation schemes ranging from non-resonant continuous-wave excitation to phonon-assisted excitation and resonant two-photon excitation. The present work adds a measurement for a different combination of material and device parameters: an InAsP/InP quantum dot in a mesa under detuned quasi-resonant excitation, with no appreciable Purcell enhancement.

**Table 2.** Comparison of reported TPI results for quantum dots emitting in the telecom C-band. The table lists the material system, growth method, photonic structure (device), excitation scheme, $g^{(2)}(0)$, and TPI visibility (V). SK denotes Stranski–Krastanov growth, DE – droplet epitaxy, MB – metamorphic buffer, WL – wetting layer, and CBG – circular Bragg grating. LA denotes longitudinal acoustic phonon-assisted excitation, TPE – two-photon excitation, RF – resonant fluorescence, and CW – continuous-wave excitation. Entries marked CW report peak visibilities. All other entries correspond to pulsed excitation and report time-integrated visibilities.

| Material | Growth | Device | Excitation | $g^{(2)}(0)$ | $V$ | Refs. |
|---|---|---|---|---|---|---|
| InAsP/InP | MBE, SK | mesa | Quasi resonant | 7.6% | 10.8% | This work |
| InAs/InAlGaAs | MBE, SK | CBG | LA | 1.7% | 91.7% | [12] |
| InAs/InAlGaAs | MBE, SK | CBG | TPE | 0.7% | 90% | [32] |
| InAs/InAlGaAs | MBE, SK | CBG | Quasi-resonant | 3.07% | 71.9% | [26] |
| InAs/InGaAs (MB) | MOVPE, SK | CBG | LA | 6.9% | 34.9% | [21] |
| InAs/InP | MOVPE, SK | mesa | TPE | 1.5% | 29% | [25] |
| InAs/InP | MOVPE, SK | CBG | Quasi-resonant | 0.32% | 19.3% | [31] |
| InAs/InP | MOVPE, DE | planar cavity | RF, CW | - | 77% | [24] |
| InAs/InGaAs (MB) | MOVPE, SK | CBG | p-shell | 0.72% | 8.1% | [33] |
| InAs/InGaAs (MB) | MOVPE, SK | planar cavity | RF | 2.3% | 14.4% | [34] |
| InAs/InP (WL) | MOVPE, DE | planar cavity | Non-resonant, CW | 3.7% | 71.3% | [23] |

Importantly, these results were obtained without deterministic device fabrication and under detuned excitation conditions, emphasizing the properties of the quantum emitter itself. Further improvements could be achieved through photonic environment engineering to enhance the extraction efficiency beyond that provided by the existing DBR structure and increase the spontaneous emission rate via the Purcell effect. The resulting reduction of the radiative lifetime can improve photon indistinguishability by mitigating the impact of spectral diffusion and other decoherence processes if only their characteristic timescale is longer compared to the radiative lifetime. It will also decrease the overlap of responses related to consecutive excitation pulses, decreasing the uncertainty of analysis and enabling investigation of shorter pulse separations. For the biexciton–exciton cascade, the maximum indistinguishability also depends on the ratio of the two radiative lifetimes. Equal enhancement of both decay rates leaves this cascade-related limit unchanged, whereas preferential Purcell enhancement of the biexciton transition can increase it.[38]

## 3. Conclusions

In conclusion, we investigated a single as-grown InAsP/InP QD emitting in the third telecommunication window under quasi-resonant excitation. The biexcitonic state was characterized in terms of quantum optical properties. We demonstrated a low multiphoton emission probability of $\boldsymbol{g^{(2)}_{\mathrm{raw}}(0) = 0.076(6)}$ and non-zero visibility in the two-photon interference measurement. An increase in the visibility from $\boldsymbol{V = 0.038(8)}$ to $\boldsymbol{V = 0.108(18)}$ was observed when reducing the laser pulse separation from 13.1 ns to 5.3 ns. This dependence is consistent with slow spectral diffusion, although two pulse separations are insufficient to identify the dominant decoherence mechanism or determine its correlation time. Photonic engineering could improve photon indistinguishability by shortening the radiative lifetime through Purcell enhancement, with the biexciton–exciton lifetime ratio also determining the limit imposed by the radiative cascade. Furthermore, employing quasi-resonant excitation with finite energy detuning from the QD emission facilitates efficient suppression of scattered laser light, making this approach attractive for all-fiber quantum photonic technologies.

## 4. Experimental Section/Methods

### 4.1 Sample Growth and Fabrication

We investigate self-assembled InAs/InP QDs embedded directly in an InP matrix, grown via MBE in SK mode. The growth process was assisted by a ripening stage, resulting in a low nanostructure density of $2\times10^9$ cm$^{-2}$ and improving in-plane shape symmetry. Due to the intermixing of P and As atoms during epitaxy, the effective composition of the nanostructures is InAsP. To enhance the photon extraction efficiency, a distributed Bragg reflector (DBR) was integrated beneath the QD layer. The QDs are located within non-deterministically fabricated mesa structures with a diameter of ~1 μm, which, combined with the DBR, provided an extraction efficiency of approximately 13.3% into a numerical aperture of 0.4 in our previous study.[27]

### 4.2 Optical Characterization

For steady-state characterization, including power- and polarization-dependent PL, the sample was mounted in a continuous-flow helium cryostat (Janis ST500) and maintained at 5 K. Excitation was provided by a continuous-wave (cw) laser at 640 nm. For PLE mapping, a tunable semiconductor cw IR laser was utilized. Both excitation and emission collection were performed using an infinity-corrected microscope objective (NA = 0.4, x 20 magnification), yielding a diffraction-limited laser spot of approximately 1 μm for the visible laser light and 3 μm for the IR laser on the sample surface. To ensure high spectral purity and eliminate spontaneous emission background or laser line tails, the IR excitation beam was filtered by a monochromator, resulting in an effective laser linewidth of 0.24 meV.

### 4.3 Quantum Optical Spectroscopy

Time-resolved correlation measurements were performed using a mode-locked Ti:Sapphire laser (Coherent Mira-HP) generating ultra-short pulses (~1.5 ps duration) at a 76 MHz repetition rate. An optical parametric oscillator was employed to extend the spectral reach into the infrared range, enabling quasi-resonant excitation. The excitation pulses were spectrally cleaned using free-space bandpass and shortpass filters.

Two-photon interference (TPI) was realized using a hybrid fiber-based/free-space setup: while the interference occurred in a fiber-based 50:50 beam splitter, the temporal delay (for both Hong-Ou-Mandel and Mach-Zehnder configurations) and polarization compensation were controlled in free space to maintain high visibility. The detection system comprised two superconducting nanowire single-photon detectors, with system efficiencies exceeding 80% in the third telecom window. The temporal resolution of the coincidence electronics was set to 100 ps (time-bin width).

**Acknowledgements**

This work was co-financed by the Ministry of Education and Science, Republic of Poland within the “Perły Nauki” project, grant no. PN/01/0117/2022. This work was also financially supported by the Bundesministerium für Forschung, Technologie und Raumfahrt-BMFTR in the frame of the project QR.N (16KIS2204), the DFG-Heisenberg grant (505496601), and the DFG project (430159793). We acknowledge Andrei Kors for his assistance in the MBE growth process, Kerstin Fuchs and Dirk Albert for technical support.

**Conflict of Interest**

The authors have no conflicts to disclose.

**Supporting Information**

Supporting Information is available from the Wiley Online Library or from the author.

A 32 meV excitation-emission detuning enables efficient spectral filtering of the excitation laser from telecom-band single-photon emission in MBE-grown InAsP/InP quantum dots. Despite the large detuning, Hong–Ou–Mandel measurements reveal non-zero photon indistinguishability, with the visibility increasing from $\boldsymbol{V_{\mathrm{fit}} = 0.544(63)}$ to $\boldsymbol{V_{\mathrm{fit}} = 0.70(10)}$ as the excitation pulse separation is reduced.

**Two-photon interference from as-grown InAsP/InP quantum dots under detuned excitation**

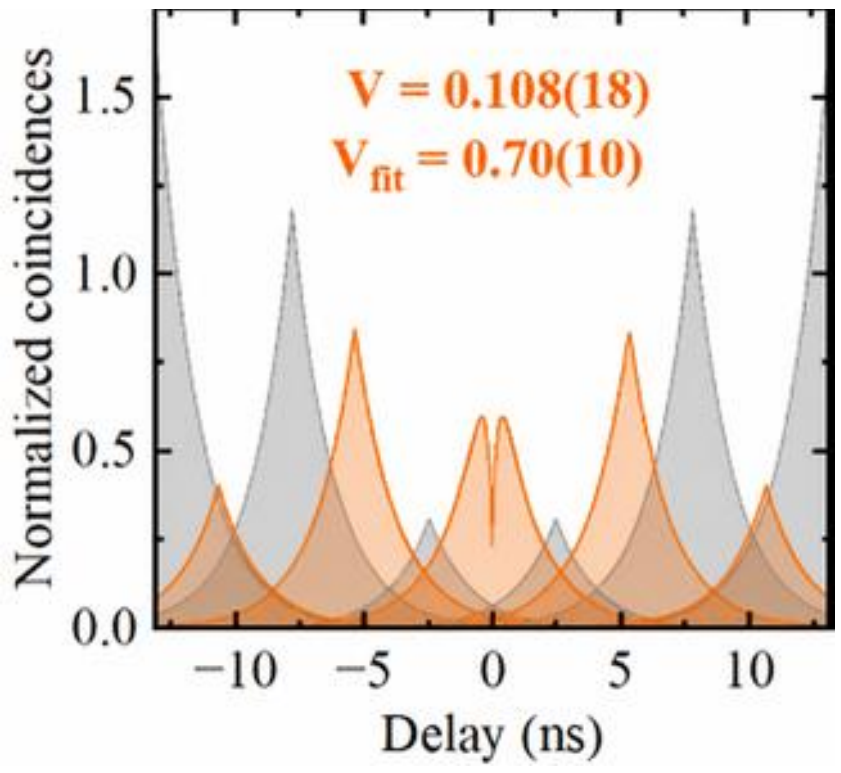

# Supporting Information

**Supporting Information: Two-photon interference from as-grown InAsP/InP quantum dots under detuned excitation**

*Maja Wasiluk*, Anna Musiał, Paweł Mrowiński, Johann Peter Reithmaier, Mohamed Benyoucef, Wojciech Rudno-Rudziński*

This Supporting Information provides additional analysis of the source lifetime to explain the discrepancies between the decay times $\tau_{dec}$ extracted from the different experiments presented in the main manuscript, as well as justification for the method of evaluation of the TPI visibility for the 5.3 ns delay.

**Source lifetime**

**Figure S1** shows the results of the time-resolved photoluminescence measurements using time-correlated single-photon counting technique. A linear fit was performed for the 3–4 ns time range on the logarithmic plot shown in Figure S1(b). The inverse of the fitted slope was taken as the effective lifetime and plotted as a function of excitation power in Figure S1(c). All the values are higher than the $\tau_{dec}$ obtained from fitting the histograms in the main manuscript, even for high excitation powers.

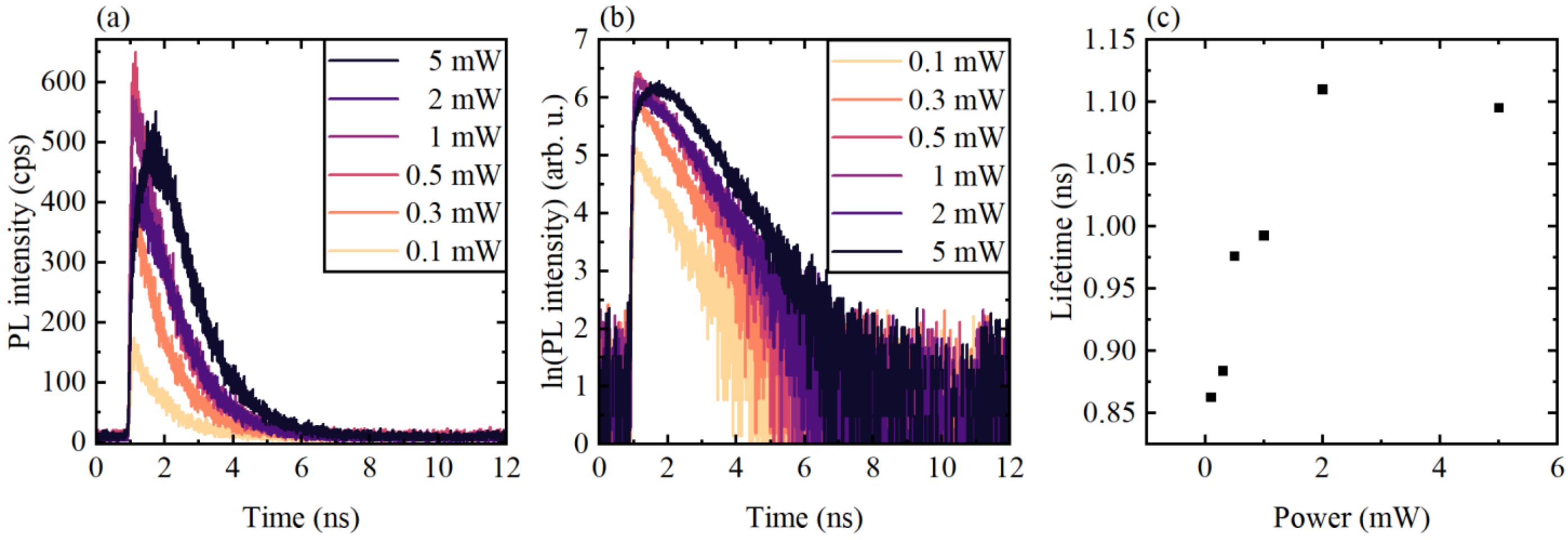

**Figure S1.** Time-resolved photoluminescence measurements. (a) PL decay curves recorded for different excitation powers. (b) Natural logarithm of the PL intensity as a function of time. (c) Extracted lifetimes as a function of excitation power.

Here, the fitting was performed only for the narrow range of time, where the data exhibit linear dependence for all the excitation powers, while the histograms were fitted by monoexponential decay for the whole range of the peak. To check the impact of the fitting range on the extracted decay time, we performed fitting of the data for 1 mW excitation power for different time ranges. **Figure S2** shows the data with 3 fitting curves on the logarithmic scale to enhance the fit discrepancy from the measured data for longer time scales.

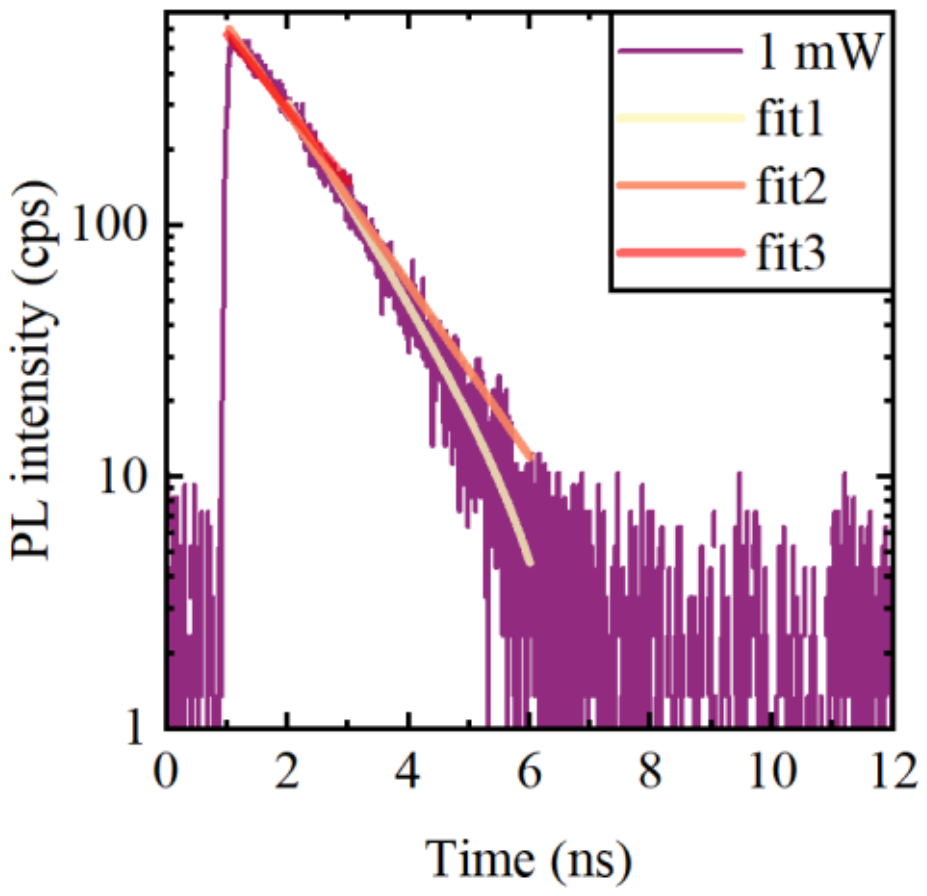

**Figure S2.** PL decay data for the 1 mW excitation power with three fitting curves of the mono-exponential decay for the time range of 2-6 ns (fit1), 1-6 ns (fit2), and 1-3 ns (fit3).

The fitting ranges were chosen as 2–6 ns, 1–6 ns, and 1–3 ns, giving decay times of 1.1 ns, 1.3 ns, and 1.5 ns, respectively. These values are in good agreement with the decay times extracted from the HBT and HOM histograms for the 13.1 ns and 5.3 ns pulse separations, respectively. The longer decay time obtained for the 5.3 ns pulse separation is attributed to the stronger overlap between neighboring peaks, which makes the short-time region near the peak maximum dominate the fitting procedure.

Finally, the low uncertainties of $\tau_{\mathrm{dec}}$ reported in the main manuscript originate solely from the fitting procedure and therefore reflect only the statistical uncertainty of the fit parameters. They do not account for the finite temporal resolution of the correlation histograms (100 ps bin width), which introduces an additional source of uncertainty.

**TPI visibility**

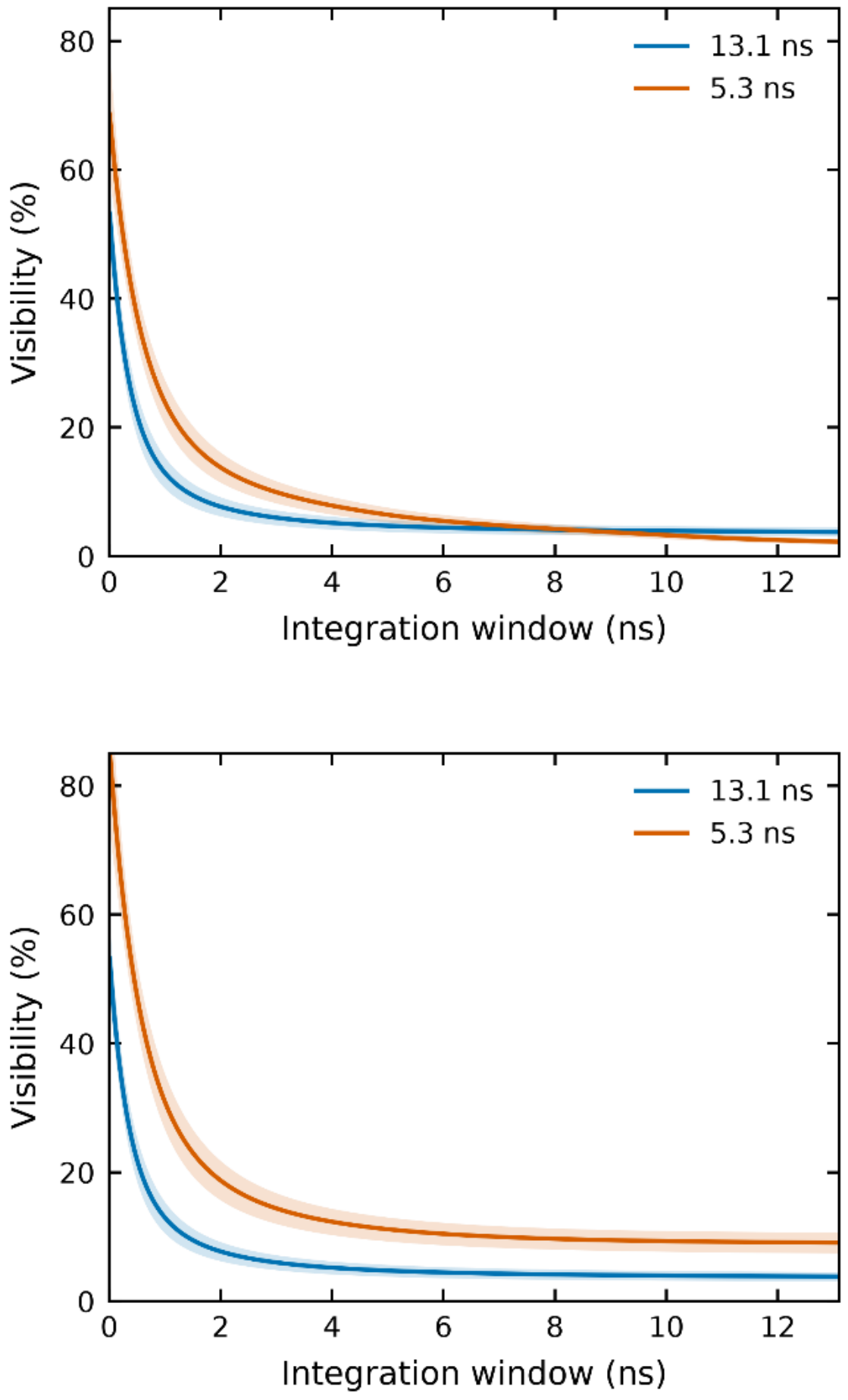

**Figure S3.** Two-photon interference visibility as a function of integration-window width for pulse separations of 13.1 ns and 5.3 ns. The upper panel includes all overlapping histogram peaks; the lower panel includes only the isolated central peak. Background is included in both evaluations. Shaded bands indicate uncertainties.

The visibility is evaluated by comparing the integrated fitted coincidence function with the distinguishable-photon reference generated using the same parameters with the interference contribution switched off. For a pulse separation of 5.3 ns and an integration window of 5.3 ns, the isolated central contribution gives V = 0.108(18), whereas the full histogram gives V = 0.061(11). These values refer to this finite integration window, not to the large-window asymptotes. Widening the integration window adds contributions from neighbouring peaks to the full-histogram integral, reducing the extracted visibility, which can fall below that obtained at a pulse separation of 13.1 ns. This dependence on integration-window width should not be confused with saturation as a function of pulse separation.